\documentclass[journal]{IEEEtran}

\usepackage{cite}
\usepackage{graphicx}
\usepackage{colortbl}
\usepackage{amssymb}
\usepackage{gensymb}
\usepackage{tabularx}
\usepackage{float}
\usepackage{amsthm}
\usepackage{amsmath}
\usepackage[switch]{lineno}
\usepackage{array}
\usepackage{mathtools}
\usepackage{float}
\usepackage{makecell}

\begin{document}

\title{Upper Esophageal Sphincter Opening Segmentation with Convolutional Recurrent Neural Networks in High Resolution Cervical Auscultation}

\author{Yassin Khalifa,
	    Cara Donohue,
        James L. Coyle,
        and~Ervin Sejdi\'{c},~\IEEEmembership{Senior Member,~IEEE}
\thanks{Manuscript received December 8, 2019; revised May 5, 2020; accepted May 30, 2020. This work was supported by the Eunice Kennedy Shriver National Institute of Child Health \& Human Development of the National Institutes of Health under Award Number R01HD092239, while the data was collected under Award Number R01HD074819. \textit{(Corresponding author: Ervin Sejdi\'{c}.)}}
\thanks{Yassin Khalifa is with the Department of Electrical and Computer Engineering, Swanson School of Engineering, University of Pittsburgh, Pittsburgh, PA 15260 USA (e-mail: yassin.khalifa@pitt.edu).}
\thanks{Cara Donohue and James L. Coyle are with the Department of Communication Science and Disorders, School of Health and Rehabilitation Sciences, University of Pittsburgh, Pittsburgh, PA 15260 USA (e-mail: cad191@pitt.edu; jcoyle@pitt.edu).}
\thanks{Ervin Sejdi\'{c} is with the  Department of Electrical and Computer Engineering, Swanson School of Engineering, University of Pittsburgh, Pittsburgh, PA 15260 USA, with the Department of Bioengineering, Swanson School of Engineering, University of Pittsburgh, Pittsburgh, PA 15260 USA, with the Department of Biomedical Informatics, School of Medicine, University of Pittsburgh, Pittsburgh, PA 15260 USA, and also with Intelligent Systems Program, School of Computing and Information, University of Pittsburgh, Pittsburgh, PA 15260 USA (e-mail: esejdic@ieee.org).}
\thanks{Digital Object Identifier 10.1109/JBHI.2020.3000057}}

\markboth{IEEE Journal of Biomedical and Health Informatics}%
{Khalifa \MakeLowercase{\textit{et al.}}: Upper Esophageal Sphincter Opening segmentation with Convolutional Recurrent Neural Networks in High Resolution Cervical Auscultation}

\IEEEtitleabstractindextext{%
\begin{abstract}
Upper esophageal sphincter is an important anatomical landmark of the swallowing process commonly observed through the kinematic analysis of radiographic examinations that are vulnerable to subjectivity and clinical feasibility issues. Acting as the doorway of esophagus, upper esophageal sphincter allows the transition of ingested materials from pharyngeal into esophageal stages of swallowing and a reduced duration of opening can lead to penetration/aspiration and/or pharyngeal residue. Therefore, in this study we consider a non-invasive high resolution cervical auscultation-based screening tool to approximate the human ratings of upper esophageal sphincter opening and closure. Swallows were collected from 116 patients and a deep neural network was trained to produce a mask that demarcates the duration of upper esophageal sphincter opening. The proposed method achieved more than 90\% accuracy and similar values of sensitivity and specificity when compared to human ratings even when tested over swallows from an independent clinical experiment. Moreover, the predicted opening and closure moments surprisingly fell within an inter-human comparable error of their human rated counterparts which demonstrates the clinical significance of high resolution cervical auscultation in replacing ionizing radiation-based evaluation of swallowing kinematics.
\end{abstract}

}

\maketitle

\IEEEdisplaynontitleabstractindextext

\IEEEpeerreviewmaketitle

\section{Introduction}
\label{S:Introduction}
	\IEEEPARstart{S}{wallowing} is a  complex process  that involves the coordination of various anatomical structures, muscles, and the biomechanical events they perform, in a somewhat sequential order to safely and efficiently transport food and liquids from the oral cavity to the stomach \cite{miller_neurobiology_2008, bhattacharyya_prevalence_2014}. Because swallowing requires the coordination of multiple subsystems of the body, a variety of medical or surgically related conditions can cause  swallowing impairments, also known as dysphagia \cite{bhattacharyya_prevalence_2014, murray_manual_1998, lazarus_swallowing_1987}. Dysphagia is prevalent with approximately 16\%-22\% of people over the age of 50, 12\%-13\% of short-term  care patients, and up to 60\% of nursing home residents experiencing swallowing difficulties \cite{cook_aga_1999, lindgren_prevalence_1991, siebens_correlates_1986}. Dysphagia can result in aspiration, or the entry of food and/or liquid into the airway below the level of the vocal folds. Aspiration of food and liquids is concerning, especially silent aspiration (Asymptomatic), because it can lead to adverse outcomes including pneumonia, malnutrition, and dehydration \cite{siebens_correlates_1986, martin-harris_videofluorographic_2008, ishida_hyoid_2002,yu_silent_2019}, as well as reduced quality of life \cite{plowman-prine_relationship_2009, miller_hard_2006, lun_chow_prevalence_2004, farri_social_2007, gustafsson_dysphagia_1991, ekberg_social_2002, perry_dysphagia_2001, nguyen_impact_2005}.\par
	
	Among the most important physiologic correlates of healthy swallowing function is the duration of upper esophageal sphincter (UES) opening (DUESO).   UES opening enables food and liquid to enter the esophagus \cite{singh_upper_2005, ahuja_assessing_2016, kahrilas_upper_1988, cook_opening_1989}. Reduced UES opening diameter, delayed onset of opening, or premature closure attenuate DUESO and can result in pharyngeal residue that in turn can enter the upper (laryngeal penetration) or lower (aspiration) airway, which are known risk factors for pneumonia and airway obstruction \cite{kim_upper_2015}. UES opening is the product of hyolaryngeal excursion, bolus propulsion, and neural inhibitory relaxation of the UES itself \cite{kim_upper_2015, cook_opening_1989}. UES dysfunction may occur due to neurological diseases that alter the timing of UES relaxation and the delivery of muscular traction forces that act to distend the relaxed UES during swallowing, or due to impaired propulsive forces applied by the oropharyngeal pump \cite{kim_upper_2015,ahuja_assessing_2016}.
	
	\begin{table}[!h]
		\caption{Summary of tools used for diagnostic assessment of UES.}
		\centering
		\begin{scriptsize}
			\begin{tabular}{p{0.15\columnwidth}|p{0.35\columnwidth}|p{0.35\columnwidth}}
				\hline
				\hline
				Modality& Strengths &Weaknesses\\
				\hline
				VFSS\cite{ahuja_assessing_2016}&- Dynamically visualize UES during all phases of swallowing\newline- Provides the exact moments when UES opens and closes&- Subjective interpretation\newline- Radiation exposure\\
				\hline
				FEES\cite{merati_-office_2013}&- Direct visualization of swallowing pharyngeal stage&- Limited in describing UES activity (either probe is covered with bolus or already through UES)\\
				\hline
				CT/MRI\cite{ahuja_assessing_2016}&- Panoramic and full-thickness visualization of oropharyngeal structures&- Hard to conduct\newline- Radiation exposure (CT)\newline- Require synchronization with patient behavior (MRI)\newline- Limited availability\\
				\hline
				Manometry\cite{ahuja_assessing_2016}&- Monitor UES pressure during swallowing\newline- Detect UES impaired relaxation/distension &- Invasive\newline- Subjective interpretation\newline- Limited availability\\
				\hline
				\hline
				EMG\cite{ahuja_assessing_2016}&- Monitor muscle activations during swallowing\newline- Detect UES impaired relaxation/distension &- Can't tell the exact moments when UES opens/closes\newline- Subjective interpretation\\
				\hline
				\hline
			\end{tabular}
		\end{scriptsize}
		\label{tab:imagmod}
	\end{table}
	
	Table \ref{tab:imagmod} summarizes the different diagnostic modalities that can generate images and signals for the assessment of UES function \cite{ahuja_assessing_2016, butler_reliability_2015, kelly_assessing_2007}. The modalities include videofluoroscopic swallow studies (VFSSs), fast pharyngeal CT/MRI, fiberoptic endoscopic evaluation of swallowing (FEES), and non-imaging instrumental tests such as pharyngeal manometry and Electromyography (EMG). Most of these modalities require expertise to perform and highly trained clinicians to interpret. VFSSs are most frequently and actually the best modality to clinically assess swallow kinematic events such as UES opening, because of the ability to dynamically visualize the UES during all phases of the swallow and give exact estimates of the moments when UES opens and closes \cite{ahuja_assessing_2016, singh_upper_2005}. However, VFSSs, which use ionizing radiation to produce radiographic images with full temporal resolution, are unavailable or undesirable to many patients, are relatively expensive, and require specialized instrumentation and trained clinicians to perform and interpret, leaving many patients undiagnosed or inaccurately diagnosed, and exposed to ongoing risk of dysphagia-related complications \cite{singh_upper_2005}.\par
	
	The holy grail of dysphagia clinical evaluation methods has long been a noninvasive and clinically feasible method of accurately identifying the biomechanical events of swallowing that contribute to airway protection such as UES opening.  The availability of such methods would enable the development of a screening tool that can differentiate between impaired and healthy swallowing with a high degree of sensitivity and specificity without the uncertainty of clinical examinations or the lack of availability of imaging studies \cite{cook_opening_1989, cook_timing_1989, daggett_laryngeal_2006, kim_temporal_2005, gross_coordination_2008}. To address the obstacle of insufficient access to instrumental testing of swallowing function universally, high resolution cervical auscultation (HRCA) is currently being investigated as an affordable, feasible, non-invasive bedside assessment tool for dysphagia. HRCA combines the use of vibratory signals from an accelerometer with acoustic signals from a microphone attached to the anterior neck region during swallowing. Following collection of signals, advanced machine learning techniques are used to examine the association between HRCA signals and physiological events that occur during swallowing \cite{sejdic_computational_2019, mao_neck_2019}. 
	
	HRCA has shown strong associations with multiple factors that affect the UES opening process. For instance, HRCA has been used in multiple studies to monitor the pharyngeal bolus propulsion during swallowing from the moment the bolus passes the mandible till the UES closes \cite{sejdic_segmentation_2009,damouras_online_2010,lee_time_2008,khalifa_non-invasive_2020}. Furthermore, hyolaryngeal excursion has been investigated to be the origin of HRCA signals in many occasions \cite{zoratto_hyolaryngeal_2010,rebrion_high-resolution_2019,he_association_2019}, and later they were successfully used to actually track the location of the hyoid bone during swallowing \cite{mao_neck_2019}. The formerly mentioned events are all parts of the UES opening mechanism which proves the potential of HRCA signals in detection of UES opening. While previous studies have monitored changes in HRCA signal features at the moments of UES opening and closure \cite{kurosu_detection_2019}, no studies have used HRCA signals to measure the time of UES opening and closure within a swallow.\par
	
	As mentioned previously, UES opening is the result of a mechanism that is controlled by multiple events occurring during swallowing, which necessitates the temporal modeling of the whole swallow for the purpose of UES opening detection. Recurrent neural networks (RNNs) have been extensively employed for the time series modeling in the recent years, due to their capability of carrying information from arbitrarily long contexts, selective information transfer across time steps, and affordable scalability when compared to stochastic models and feed-forward networks \cite{lipton_critical_2015,elman_finding_1990,ravi_deep_2017,khalifa_sparse_2017}. RNNs are seemingly efficient in modeling temporal contexts in time series data and have been recently used to perform many biomedical prediction tasks of various complexities \cite{chauhan_anomaly_2015,spampinato_deep_2017,sano_multimodal_2019,cui_identifying_2019}, but nevertheless using RNNs on raw signals is extremely hard to optimize because of the propagation of error signals through huge number of time steps \cite{stollenga_parallel_2015,lecun_deep_2015}. To overcome this, convolutional neural networks (CNNs) have been utilized for the perception of short contexts and more abstraction before feeding into RNNs for the perception of longer temporal contexts \cite{stollenga_parallel_2015}. Known as representation learning, such hybrid architectures allow feeding the machines with raw data to automatically discover representations necessary for the detection problem \cite{lecun_deep_2015}. These models were first conceived for computer vision applications \cite{stollenga_parallel_2015,krizhevsky_imagenet_2012}; however, similar designs are being adopted recently for event detection in biomedical signals \cite{tan_application_2018,xiong_ecg_2018} in addition to numerous applications in audio and speech signal processing \cite{adavanne_sound_2019}.
	
	In this study, we propose a novel deep learning approach that uses HRCA acceleration signals to estimate the moments at which the UES opens and re-closes during swallowing and compare the estimates to gold-standard judgments of DUESO in videofluoroscopic images. The proposed method relies on convolutional recurrent neural networks to extract the dynamics of the swallowing vibrations from HRCA signals and use them to infer the moments when the UES first opens and re-closes during swallowing. Verifying the ability of HRCA signals to demarcate the UES opening among other swallowing physiological events, will promote a new non-invasive sensor-based swallowing assessment technology that is widely available and doesn't add financial or relocation burdens to patients. Moreover, it will help patients get a consistent feedback about their swallowing, while they are swallowing; a feature that will not only help improve the clinic-based swallowing evaluation, but will also be of a great benefit for the patients towards feeling the progress of the rehabilitation process and maintaining safe swallowing.\par

\section{Methodology}
\label{S:Methodology}
	\subsection{Materials and Methods}
	\label{SS:MaterialsnMethods}

		Permission for this study was granted by the institutional review board of the University of Pittsburgh and all participating patients provided informed consents including consent to publish before enrollment. A total of one hundred and sixteen patients (72 males, 44 females, age: $62.7 \pm 15.5$) with suspected dysphagia resulting from a variety of diagnoses, underwent an oropharyngeal swallowing function evaluation by a speech language pathologist using VFSS at the University of Pittsburgh Medical Center Presbyterian Hospital (Pittsburgh, PA). Of the sample, 15 patients were diagnosed with stroke while the remaining 101 patients were diagnosed with different medical conditions unrelated to stroke.\par
		
		Swallows for this study, were collected as a part of standard clinical care rather than for research purposes alone. As a result, speech language pathologists who conducted the VFSSs, had the ability to alter the evaluation protocol based on the patient's clinical manifestation of dysphagia. This included how the boluses were administered to patients (i.e. spoon, cup), the volume and viscosity/texture of each bolus of food and liquids, the number of trials, and head position during swallowing (i.e. head/neck flexion, head rotation, head neutral). The following consistencies were used during VFSSs: thin liquid (Varibar thin, Bracco Diagnostics, Inc., $< 5$ cPs viscosity), mildly thick liquid (Varibar nectar, $300$ cPs viscosity), puree (Varibar pudding, $5000$ cPs viscosity), and Keebler Sandies Mini Simply Shortbread Cookies (Kellogg Sales Company). Boluses were either self-administered by patients via a cup or a straw or administered by the clinician through the use of a spoon ($3-5$ mL).\par
		
		This study yielded 710 swallows (132 from patients diagnosed with stroke and 578 from patients with other diagnoses) with an average duration of pharyngeal bolus transit of $869.5 \pm 221$ msec and an average DUESO of $604.9 \pm 150$ msec. The collected swallows were classified into three categories: single (single bolus swallowed with one swallow), multiple (single bolus swallowed using more than one swallow), or sequential (multiple boluses swallowed sequentially in a rapid manner). The final data included 224 single, 477 multiple, and 9 sequential swallows.\par	
		
		\begin{figure}[t]
			\centering
			\includegraphics[width=\columnwidth]{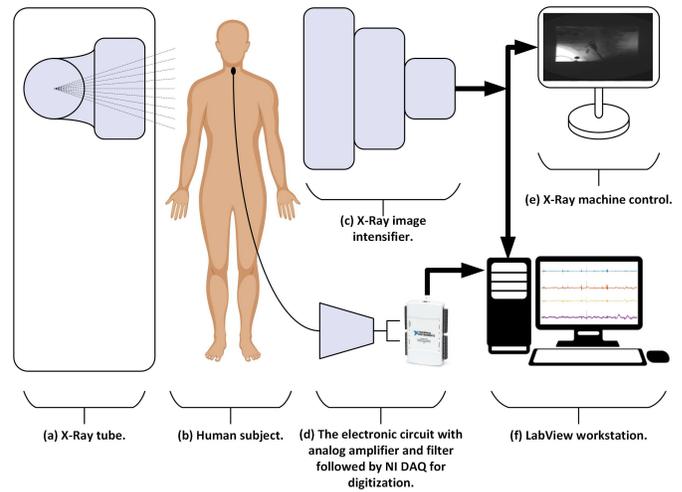}\\
			\caption{The experimental setup of the study. (a) An X-Ray tube that resides in a table is adjusted in a vertical position to be parallel to the swallowing path. (b) The human subject is standing or comfortably seated between the x-ray tube and the image intensifier with the HRCA sensors attached to the anterior neck. (c) The image intensifier is positioned and adjusted according to the subject height, so that the produced frames capture all of the important anatomical landmarks of the oropharyngeal swallow (jaws, pharynx, and esophagus). (d) The sensors are connected to the electronic circuit that supplies power and performs analog amplification and filtration and then to the NI DAQ for sampling. (e) The video feed is taken directly from the image intensifier to the X-Ray control workstation where clinicians and radiologists create, save , and view the exams. (f) The video feed from the image intensifier is cloned into the video capture card installed on the research workstation which is also connected to the NI DAQ and runs LabView for means of data collection and synchronization.}
			\label{fig:expsetup}
		\end{figure}
		
	\subsection{Data Acquisition}
	\label{SS:Data Acquisition}
		The general experimental setup is illustrated in Fig. \ref{fig:expsetup}. During all recording sessions, VF equipment was controlled by a radiologist and the patients were comfortably seated with the swallowing sensors attached to the anterior neck region using double sided tape. VF was conducted in the lateral plane using a Precision 500D system (GE Healthcare, LLC, Waukesha, WI) at a pulse rate of 30 pulses per second (PPS) and with the images acquired at a frame rate of 30 frames per second (FPS) \cite{bonilha_preliminary_2013}. The video stream was captured and digitized using an AccuStream Express HD video card (Foresight Imaging, Chelmsford, MA) into movie clips with a resolution of $720\times 1080$ at 60 FPS. \par
		
		A tri-axial accelerometer (ADXL 327, Analog	Devices, Norwood, Massachusetts) and a contact microphone (model C 411L, AKG, Vienna, Austria) were used to collect swallowing vibratory and acoustic signals. The accelerometer was mounted into  a small plastic case with a concave surface that fits on neck curvature and the case was attached to the skin overlying the cricoid cartilage using a tape. The accelerometer was attached such that its main axes are aligned parallel to the cervical spine, perpendicular to the coronal plane, and parallel to the axial/transverse plane. These axes are referred to as superior-inferior (S-I), anterior-posterior (A-P), and medial-lateral (M-L) respectively. The microphone was mounted towards the right lateral side of the larynx to avoid contact noise with the accelerometer and guarantee a clear radiographic view of the upper airway. Attaching the sensors around the area of cricoid cartilage is logical given that most of the pharyngeal swallowing activity is produced by the anatomical structures present at this level and it has been reported to yield the best signal-to-noise ratio for the acquisition of swallowing signals \cite{cichero_physiologic_1998,lee_time_2008,dudik_dysphagia_2015,khalifa_non-invasive_2020}. 
		
		The accelerometer has a bandwidth of 1600 Hz after which the response falls to -3dB of the response to low frequency acceleration. In other words, the accelerometer has a low pass filter with a cut-off frequency at 1600 Hz. The contact microphone was chosen as well so that it produces a flat frequency response over the entire range of audible sounds which was proved to pass most of the frequencies encountered during swallowing \cite{takahashi_methodology_1994,cichero_acoustic_2002,dudik_dysphagia_2015}. The signals from both the accelerometer and microphone were hardware band-limited to 0.1-3000 Hz with an amplification gain of 10. The cut-off frequencies for the band-limiting filter were chosen so that most of body sway components below 0.2 Hz are suppressed and the signal components with the vast majority of energy are passed \cite{lee_time_2008,cichero_acoustic_2002,lee_effects_2010,el-jaroudi_application_1996}. The signals were sampled using a National Instruments 6210 DAQ at a sampling rate of 20 kHz. Both signals and video were acquired simultaneously using LabView’s Signal Express (National Instruments, Austin, Texas) with a complete end-to-end synchronization.
		
	\subsection{VF Image Analysis}
	\label{SS:VFSS Analysis}
		Video clips were segmented based on individual swallow events by tracking the bolus in a frame by frame manner. The onset of the pharyngeal swallow event was defined as the frame in which the head of the bolus passes the shadow of the posterior border of the ramus of the mandible and the offset as the frame in which the bolus tail passes through the UES \cite{lof_test-retest_1990}, in order to capture the entire duration of pharyngeal bolus flow. Three expert judges trained in swallow kinematic judgments, identified the video frame of first UES opening and the video frame of first UES closure in the segmented videos. All raters who segmented swallowing videos and analyzed UES opening and closure established a priori intra- and inter-rater reliability with ICC's over 0.99. All raters maintained intra- and inter-rater reliability throughout measurements on 10\% of swallows with ICC's over 0.99 and were blinded to participant demographics and diagnosis and any bolus condition information.\par

	\subsection{Signals Preprocessing}
	\label{SS:Preprocessing}

		Numerous physiologic and kinematic events such as coughing and breathing occur in close temporal proximity to the pharyngeal swallow event. These events can contribute to the collected vibratory and acoustic signals \cite{damouras_online_2010}. As a first step to overcome confounding noise in the signals due to multi-source environmental data collection and other measurement errors, the signals accrued at a sampling rate of 20 kHz were down-sampled to 4 kHz. A more intense down-sampling could have been adopted as previous studies reported that the frequency with the maximum energy for swallowing accelerometry signals occurs below 100 Hz and the central frequency almost below 300 Hz \cite{lee_time_2008,sejdic_baseline_2010,dudik_comparative_2015,dudik_characteristics_2015}. However, we chose down-sampling to 4 kHz so that we match twice the max frequency component present in the acceleration signals (1600 Hz). Down-sampling was performed through applying an anti-aliasing low pass filter then picking up individual samples to match the new rate.\par
		
		The baseline outputs of accelerometer and microphone (produced by zero-physical input) were recorded earlier before the main data collection procedure and device noise was characterized through modified covariance auto-regressive modeling \cite{sejdic_baseline_2010,marple_new_1980}. The order of the auto-regressive model was 10 and it was determined using the Bayesian information criterion \cite{sejdic_baseline_2010}. The coefficients of the auto-regressive model were then used to create a finite impulse response filter (FIR) to remove the device noise from the recorded swallowing signals \cite{sejdic_baseline_2010}. Afterwards, the low-frequency noise components and motion artifacts were eliminated from accelerometer signals using fourth-order least-square splines \cite{sejdic_method_2012, sejdic_effects_2010}. Particularly, we used fourth-order splines with a number of knots equivalent to $\frac{N\times f_l}{f_s}$, where $N$ is the data length and $f_s$ is the sampling frequency. $f_l$ is called the lower sampling frequency and it is proportional to the frequency associated with motion artifacts. The values for $f_l$ were calculated and optimized in previous studies \cite{sejdic_method_2012}. Finally, the effect of broadband noise on signals was reduced through wavelet denoising \cite{sejdic_procedure_2010}. Specifically, we used tenth-order Meyer wavelets and soft thresholding. The threshold was calculated using $\sigma\sqrt{2\log N}$, where $N$ is the number of samples and $\sigma$ is the estimated standard deviation of the noise (calculated through down-sampling the wavelet coefficients) \cite{sejdic_procedure_2010, dudik_statistical_2016}.
		
		\begin{figure*}[!h]
			\centering
			\includegraphics[height=0.82\textheight, width=\textwidth]{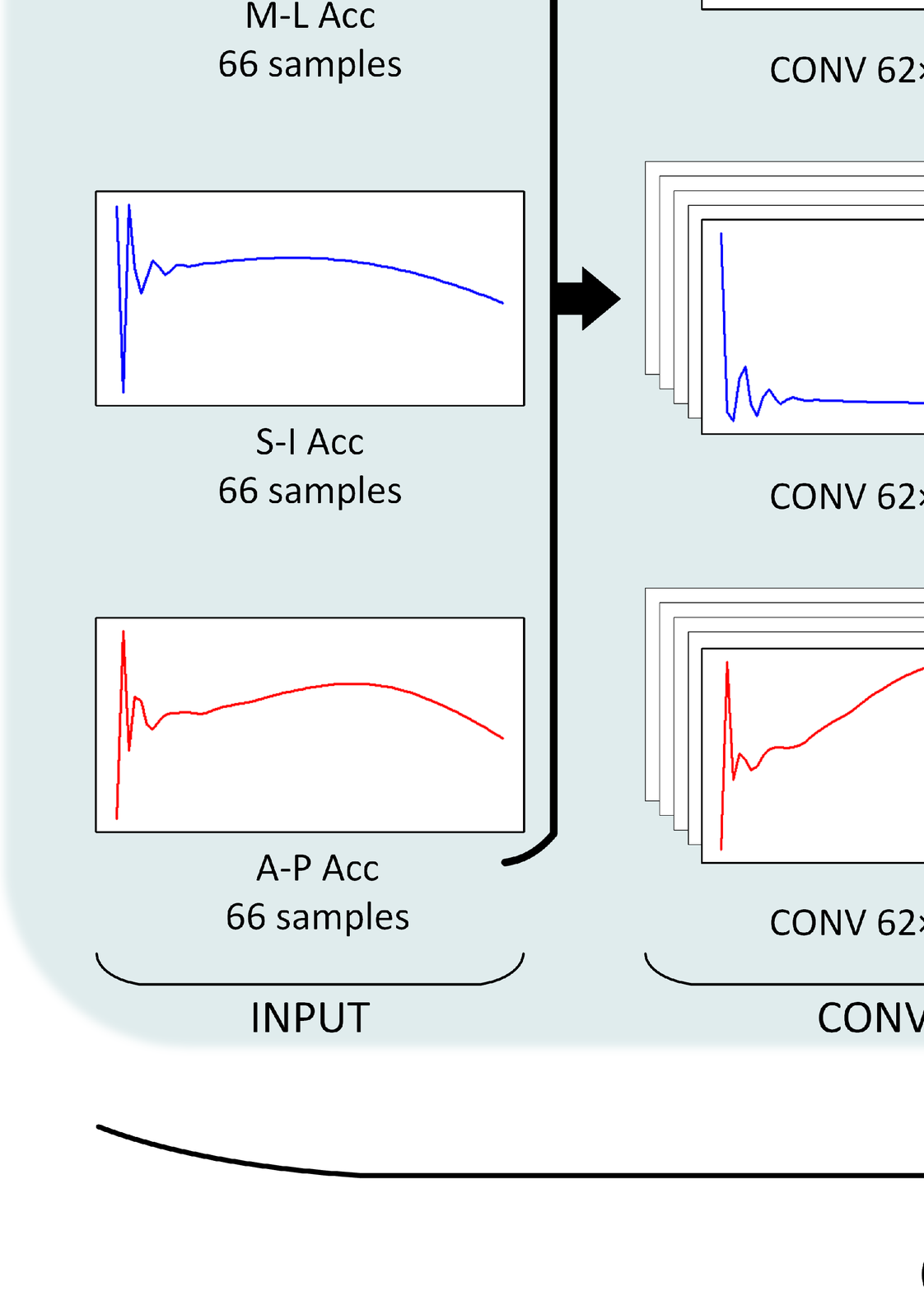}\\
			\caption{The architecture and data flow in the UES opening detection system. (a) This part is where the 3-channel acceleration signals from each swallow are denoised and split into equal chunks each of 66 samples (equivalent to 1 VF frame). (b) This part shows the operation of the CNN network part per data chunk. The architecture of the used 1D CNN which is comprised of two layers, the first applies 16 filters on each channel and produces 48 channels. The first CNN layer is followed by a max pooling layer and another CNN layer identical to the first except that it applies 1 filter per channel then a max pooling layer to reduce the size of the features.  (c) This is an illustration for the operation of the CNN after training that shows a chunk of 3-channel acceleration pushed throw the first layer of CNN to produce 16 feature-channels per original channel. The length of chunks is shorter after this layer due to convolution on the edges of the chunks (no padding is used). (d) This is an illustration that shows the architecture of the GRU unit with the reset and update parts that help propagate states across time steps. (e) ($x_{1:T}$) is the output train from the CNN for chunks ($1:T$) which is fed into the RNN units. (f) The architecture of the 3-layer RNN used for time sequence modeling. (g) The output sequence from the last layer of the RNN ($\hat{y}_{1:T}$) is flattened and fed into the first fully connected layer. (h) A diagram of the 3 fully connected layers (each of 128 units) used to combine the features coming out of the RNN. (i) The output layer of the network which is composed of 90 units ($y_{1:T}$) that resemble the UES opening mask.}
			\label{fig:sysdesign}
		\end{figure*}

\subsection{System Design}
\label{SS:System Design}
	Due to the fact that there is no specific rule of thumb to calculate the number of layers and layer sizes for a certain problem, the used architecture was fine-tuned based on an experimental approach and by following the best network configurations that achieved good results in similar problems \cite{tan_application_2018,shashikumar_detection_2018,adavanne_sound_2019}. Particularly, we tested multiple architecture depths that included more layers of CNN (3, 4, and 5 layers) with up to 32 filters per channel and more RNN unit sizes up to 128 as well as different RNN units (GRU and LSTM) with both Sigmoid and tanh recurrent activations. The chosen architecture was found to be the most stable among the tested configurations and the fastest to converge. In other words, it included the smallest number of parameters to be optimized while achieving a detection accuracy that doesn't sharply change when adding more layers or increasing the layer sizes. The used architecture employed also dropout between layers as well as early stopping techniques to control the network from over-fitting to the training data \cite{srivastava_dropout_2014}.
	
	The longest swallow event duration in the collected dataset was around 1500 msec (90 frames of VF). The signals were divided into chunks 16.67 msec in length (equivalent to one frame in VF or 66 samples in signals). Each signal chunk is composed of 3 axes of acceleration which makes the dimensions $66$ samples $\times$ $3$ channels. The chunks were fed into a 1D convolutional neural network that included two convolutional layers with a max pooling layer in between as in Fig. \ref{fig:sysdesign}. Both convolutional layers were followed by a rectified linear unit (ReLU). The first convolutional layer applied 16 "$1 \times 5$" filters per channel which produced  $3$ "$62$ features $\times$ $16$ channels". The max pooling layer applied a window of size $2$ with 2 strides and reduced the features into "$31$ features $\times$ $48$ channels". The last convolutional layer was identical to the first one except that it used only one filter per channel which produced "$27$ features $\times$ $48$ channels".\par
	
	The complete sequence of features $x_{1:T}$ (for a full swallow) coming out of the convolutional layer was then fed into a 3-layers dynamic RNN with gated recurrent units (GRUs) as building blocks each of 64 units and a sequence of 90 time steps. The RNN computed an output sequence $\hat{y}_{1:T}$ using the following nonlinear model:
	
	\begin{scriptsize}
		\begin{equation}
			\setlength\arraycolsep{1pt}
			\begin{matrix*}
			r^{(k)}_t & = & \begin{cases} 
			\sigma(W^{(1)}_r \left[h^{(1)}_{t-1}, x_{t}\right] + b^{(1)}_r), &\quad\quad\quad\text{k=1,}\\
			&\\
			\sigma(W^{(k)}_r \left[h^{(k)}_{t-1}, h^{(k-1)}_{t}\right] + b^{(k)}_r), &\quad\quad\quad\text{k=2, 3}
			\end{cases}\nonumber\\
			&&\nonumber \\
			z^{(k)}_t & = & \begin{cases} 
			\sigma(W^{(1)}_z \left[h^{(1)}_{t-1}, x_{t}\right] + b^{(1)}_z), &\quad\quad\quad\text{k=1,}\\
			&\\
			\sigma(W^{(k)}_z \left[h^{(k)}_{t-1}, h^{(k-1)}_{t}\right] + b^{(k)}_z), &\quad\quad\quad\text{k=2, 3}
			\end{cases}\nonumber\\
			\end{matrix*}
		\end{equation}
	\end{scriptsize}
	\begin{scriptsize}
		\begin{equation}
			\setlength\arraycolsep{1pt}
			\begin{matrix*}
			\hat{h}^{(k)}_t & = & \begin{cases} 
			tanh(W^{(1)} \left[ r^{(1)}_t h^{(1)}_{t-1}, x_{t}\right] + b^{(1)}), &\text{k=1,}\\
			&\\
			tanh(W^{(k)} \left[ r^{(k)}_t h^{(k)}_{t-1}, h^{(k-1)}_{t}\right] + b^{(k)}), &\text{k=2, 3}\\
			\end{cases}\nonumber\\
			&&\nonumber \\
			h^{(k)}_{t} & = & \quad z^{(k)}_{t} \hat{h}^{(k)}_t +(1 - z^{(k)}_{t}) h^{(k)}_{t-1},\quad\quad\quad\quad\quad\quad\quad\ \text{k=1, 2, 3}\nonumber\\
			&&\nonumber \\
			\hat{y}_t & = & Uh^{(3)}_t + c \nonumber
			\end{matrix*}
		\end{equation}
	\end{scriptsize}	
	
	The output sequence $\hat{y}_{1:T}$ coming out of the RNN was masked (ones/zeros mask) before being fed in to the following stages to balance for the shorter swallows (less than 90 frames). Furthermore, the length of each swallow was considered in the architecture of the RNN and the same mask was used in the calculation of the cost function for the whole problem. The sequence was then fed in to 4 fully connected layers in order to fuse the temporal features from RNN into a meaningful UES opening segmentation mask. This part of the network featured 3-ReLU activated layers with 128 units and an output layer that assembled 90 units, one for each time step in the swallow as shown in Fig. \ref{fig:sysdesign} plus Sigmoid activation for a zeros and ones segmentation mask. Each two fully connected layers were separated by a dropout layer with a drop rate of 20\%. \par

	The final cost function was defined as the mean squared error between the zero-padded ground truth $\bar{y}_{1:T}$ labeled by the expert judges and the masked output coming from the final connected layer $\hat{y}_{1:T}$ as follows:	
	\begin{equation}
		MSE = \frac{1}{T}\sum_{i=1}^{T}\left[(\bar{y}_{i} - \hat{y}_{i})\times mask_{i}\right] ^2
	\end{equation}
	where $mask_{i}$ is the mask used to compensate for short swallows. We used the Adam optimizer to train the network due to its superiority in convergence without fine tuning for hyper-parameters \cite{bengio_practical_2012}.\par
	
	\begin{figure}[!h]
		\centering
		\includegraphics[width=\columnwidth]{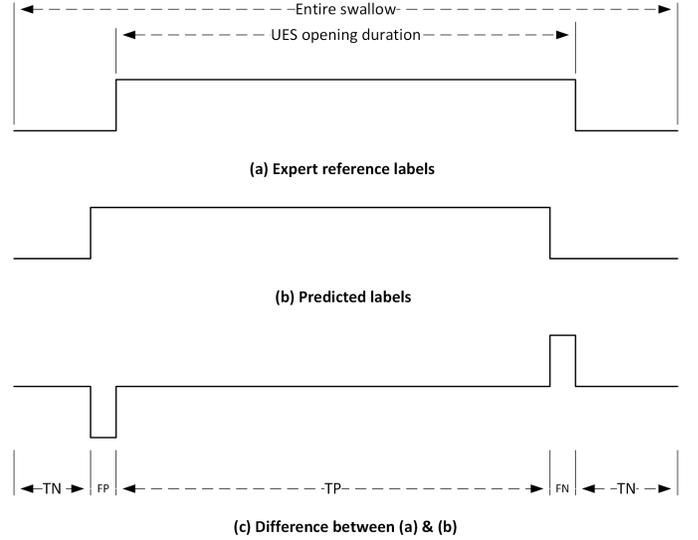}\\
		\caption{The evaluation procedure for each swallow. (a) The UES opening mask created from the expert manual segmentation in VF images. (b) The UES opening mask as predicted by the proposed algorithm. (c) Comparison is performed between the masks from (a) and (b) to create a confusion matrix. The confusion matrix is created in this way for each swallow included in testing. The values of accuracy, sensitivity, and specificity are calculated through this confusion matrix.}
		\label{fig:evalfn}
	\end{figure}
	
	\subsection{Evaluation}	
		The dataset was randomly divided into 10 equal subsets in terms of the number of swallows. A holdout method was repeated 10 times by training with 9 subsets and testing with the remaining one (10-fold cross validation). The results of the proposed system are in the form of a segmentation mask that tells when the UES opens and closes with respect to the start (onset) of the swallow segment as shown in Fig. \ref{fig:evalfn} (b). This mask is calculated for approximately each swallow in the dataset when passed as a test sample through the trained system. In order to acquire a solid evidence about the detection quality of the system, a confusion matrix is constructed for each swallow based on the predicted segmentation mask and the reference mask as labeled by judges. The confusion matrix is then used to calculate accuracy, sensitivity, and specificity as follows:
		\begin{equation*}
			\begin{matrix*}
				&Accuracy&= \frac{TP+TN}{TP+FP+TN+FN}\\
				&Sensitivity&= \frac{TP}{TP+FN}\\
				&Specificity&= \frac{TN}{FP+TN}
			\end{matrix*}
		\end{equation*}
		where TP stands for True Positive, TN stands for True Negative, FP stands for False Positive, and FN stands for False Negative. Furthermore, the difference between the actual and predicted UES opening and UES closure was measured, so that we could compare it to the human judges' tolerance reported in the literature. \par

	\subsection{Clinical Validation}
		In order to evaluate the proposed system in a clinical environment, it was tested during the workflow of an ongoing clinical experiment performed on 15 (8 males, 7 females, age: $63.7\pm 6.2$), community dwelling healthy adults who provided informed consent, and who had no reported current or prior swallowing difficulties. Participants in this validation sample also had no history of neurological disorder, surgery to the head or neck region, or chance of being pregnant based on participant's report. The experimental setup of this clinical experiment relied on the same equipment and hardware used for the collection of the main dataset as shown in Fig. \ref{fig:expsetup}. This included recording VF in the lateral plane using a Precision 500D system (GE Healthcare, LLC, Waukesha, WI) at a pulse rate of 30 pulses per second (PPS) and with the images acquired at a frame rate of 30 frames per second (FPS). The video stream was captured and digitized using an AccuStream Express HD video card (Foresight Imaging, Chelmsford, MA) at 60 FPS. Swallowing vibratory and acoustic signals were acquired concurrently with VF using the same tri-axial accelerometer and microphone (ADXL 327, Analog	Devices, Norwood, Massachusetts andmodel C 411L, AKG, Vienna, Austria). The sensors were attached to the same location on the anterior neck to the skin overlying the cricoid cartilage. The signals from both sensors were also band-limited between 0.1-3000 Hz and amplified with a gain of 10 then sampled at a rate of 20 kHz via a National Instruments 6120 DAQ through LabView's Signal Express (National Instruments, Austin, Texas).
		
		\begin{figure*}[!h]
			\centering
			\includegraphics[width=0.8\textwidth, height=0.55\textheight]{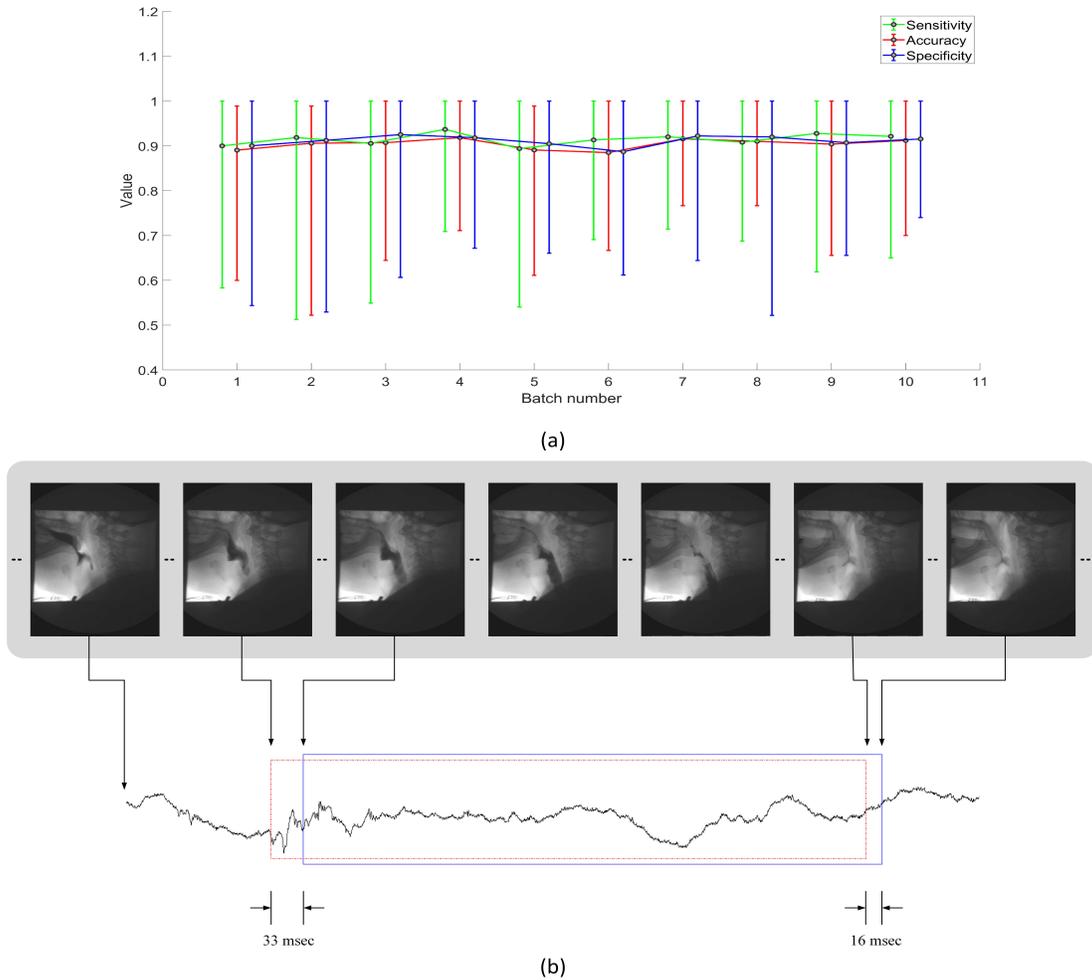}
			\caption{Distribution of per swallow based performance measurements in each testing batch of the 10-fold cross validation process and a sample visual of the detection in one of the swallows.  A sample of figures showing the timing difference between the automatically detected DUESO by our algorithm and the actual DUESO observed from VF (in frames) for both opening and closure. (a) Distribution for accuracy, sensitivity, and specificity in each batch (min, average, and max). (b) shows a sample full swallow with both the predicted (in red) and the actual DUESO (in blue) marked on the A-P acceleration component and video frames.}
			\label{fig:batch_meas}
		\end{figure*}
		
		The participants in this clinical experiment were community dwelling adults without report of current or prior swallowing difficulties. Therefore, only ten thin liquid boluses (5 at 3mL by spoon, 5 unmeasured self-selected volume cup sips) administered in a randomized order in order to limit x-ray radiation exposure. For all spoon presentations, participants were instructed by the researcher to "Hold the liquid in your mouth and wait until I tell you to swallow it." Liquid bolus presentations by cup varied in volume by participant, because participants were instructed by the researcher to "Take a comfortable sip of liquid and swallow it whenever you're ready." Fifty swallows, selected randomly from this independent clinical experiment, were used to test the system for UES opening detection after being trained over the full 710 swallows dataset.\par
	
\section{Results}
	
	A chunk of 3D acceleration ($3 \times 133$) was first preprocessed to achieve denoising and artifact removal as shown in Fig. \ref{fig:sysdesign}. After preprocessing, the filtered acceleration segments were fed into the convolutional network (CNN) part of the system as in the snapshot shown in the lower part of Fig. \ref{fig:sysdesign}. The snapshot represents a sample feature map across the CNN that shows the evolution of inputs (low-level features) into high level features at the final layer of the CNN. The later helps identify more complex features in the input signals and promote distinctive traits while the insignificant features disappear.\par
	
	\begin{table}[h]
		\caption{Summary of the performance measurements that the proposed system achieved for both the main patient and the independent clinical datasets.}
		\begin{footnotesize}
			\begin{tabular}{l|c|c}
				\hline
				\hline
				&Main dataset&Independent dataset\\
				\hline
				Average Accuracy&0.9093&0.8880\\
				\hline
				Average sensitivity&0.9145&0.8559\\
				\hline
				Average specificity&0.9119&0.9356\\
				\hline
				\makecell[l]{\% of swallows with UES\\opening error $<$ 3 VF frames}&82.6&84\\
				\hline
				\makecell[l]{\% of swallows with UES\\opening error $<$ 4 VF frames}&90&88\\
				\hline
				\makecell[l]{\% of swallows with UES\\closure error $<$ 3 VF frames}&72.3&66\\
				\hline
				\makecell[l]{\% of swallows with UES\\closure error $<$ 4 VF frames}&80&74\\
				\hline
				\hline
			\end{tabular}
		\end{footnotesize}
		\label{tab:perfmeas}
	\end{table}
	
	\begin{figure*}[h]
		\centering
		\includegraphics[width=0.85\textwidth, height=0.45\textheight]{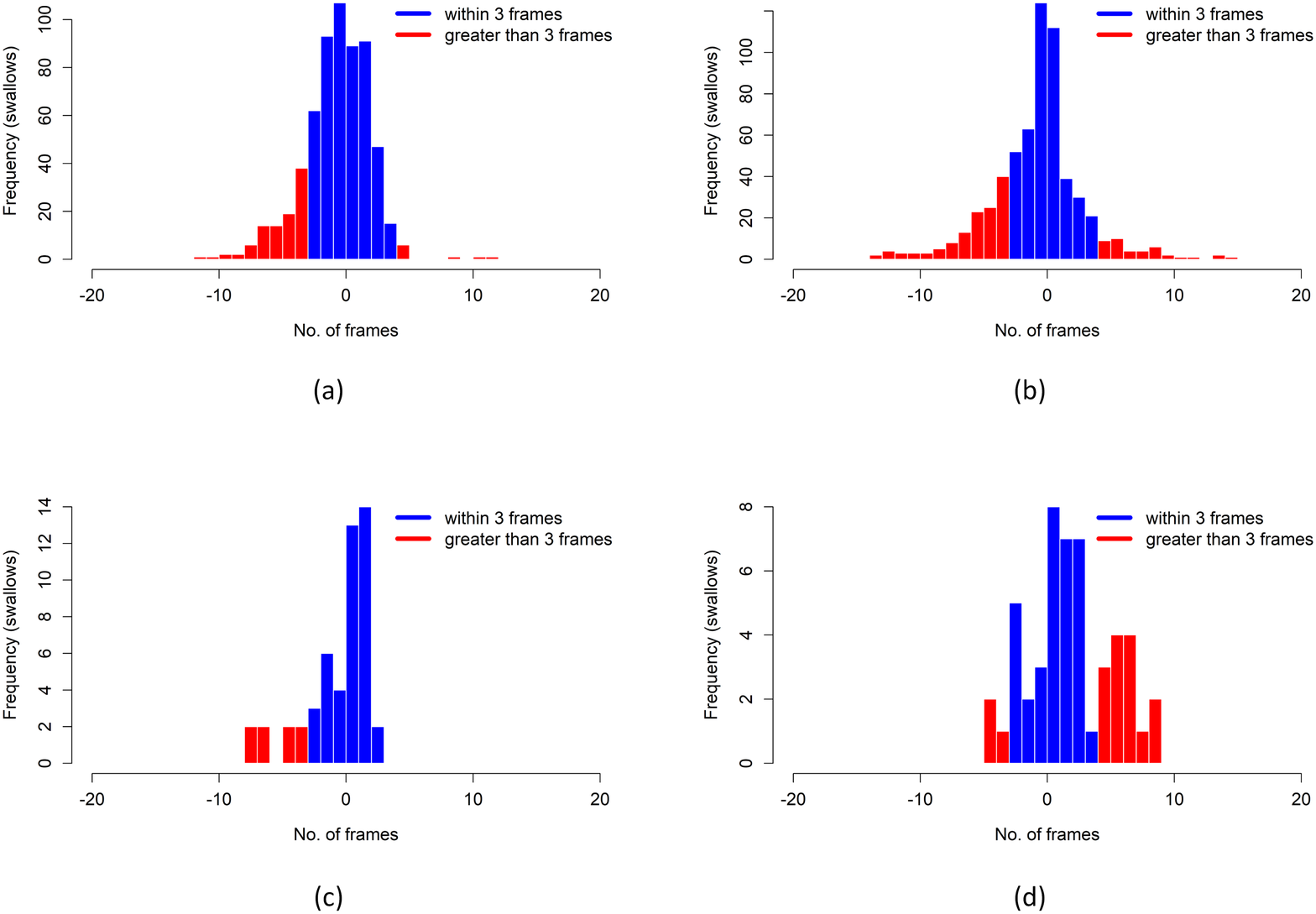}
		\caption{The timing difference between the automatically detected DUESO by the proposed system and the actual DUESO observed from VF (in frames) for both opening and closure in the whole dataset and the clinically independent data.  The differences between the detected opening frame and the opening frame marked by the judges are highlighted in (a) for the 10 folds within the original dataset and in (c) for the clinically independent data. The differences between the detected closure frame and the closure frame marked by the judges are highlighted in (b) for the 10 folds within the original dataset and in (d) for the clinically independent data. The Positive values indicate that the actual UES opening and closure preceded the predicted UES opening and closure.}
		\label{fig:doffset}
	\end{figure*}
	
	Fig. \ref{fig:batch_meas} (a) shows the performance of the proposed system across the 10-folds of the whole set of swallows. The values presented, represent the distribution of sensitivity, accuracy, and specificity in each fold. Each vertical line has 3 main points that represent the min average and maximum respectively from bottom up. The average accuracy of all folds across the whole dataset was 0.9039 with 0.9145 sensitivity and 0.9119 specificity. Fig. \ref{fig:batch_meas} (b) depicts a comparison between DUESO detection from the proposed system against the manual labeling by experts through the use of VF. On average, the network detected UES opening 33 msec earlier and closure 16 msec earlier than true opening and closure as measured by swallow kinematic analysis. The outcome of the algorithm for the whole set of swallows, was calculated and compared to the VF based labels and the differences are shown through the histograms in Fig. \ref{fig:doffset} (a-b) and Table \ref{tab:perfmeas}. The comparison shows that for 82.6\% of the swallows, the opening of UES was detected within a 100 msec ($\approx$ 3 frames at 30 FPS) of the human ratings, and within a 133 msec ($\approx$ 4 frames at 30 FPS) for 90\% of the swallows (Fig. \ref{fig:doffset} (a)). Likewise, the network accurately detected UES closure within a 100 msec ($\approx$ 3 frames at 30 FPS) for 72.3\% of the swallows and within a 133 msec ($\approx$ 4 frames at 30 FPS) for more than 80\% of the swallows (Fig. \ref{fig:doffset} (b)). The accepted tolerance for human frame selection $\approx\ \pm\ 2.48$ frames at 30 FPS \cite{lof_test-retest_1990}.\par
	
	The system also presented similar results when tested using the swallows from the independent clinical experiment as in Table \ref{tab:perfmeas}. for the 50 swallows, the system achieved an average per swallow accuracy of 0.8880, an average per swallow sensitivity of 0.8559, and an average per swallow specificity of 0.9356. Fig. \ref{fig:doffset} (c-d) show histograms for the difference between the automatic detection and the reference manual labeling of the DUESO in terms of opening and closure frames. The results showed that UES opening and closure were detected within a 100 msec tolerance in around 84\% and 66\% of the swallows in the independent test set respectively. \par
	
\section{Discussion}

	The main purpose of this study was to test the feasibility of HRCA in detecting the exact timing of UES opening and closure during swallowing using non-invasive neck-attached sensors independent of VFSS images and to compare the accuracy to human ratings of the DUESO. We have established the fact that UES opening can be best visualized using VF which is clinically impractical due to the delivered radiation doses and unavailability outside clinical care settings. We have also demonstrated the critical rule that UES plays during swallowing and how monitoring its opening and closure will help identify the risks leading to unsafe swallowing. As a necessary part of the optimal goal to create a non-invasive swallowing monitoring system, UES opening/closure detection should help patients with brainstem parts, responsible for swallowing regulation, damaged and/or surgically removed to rehabilitate and relearn how to swallow. These patients will have a consistent feedback to tell if they are correctly performing swallowing compensation maneuvers in which they are taught to improve the hyolaryngeal excursion which would in turn reflect on UES duration/diameter and airway protection in order to maintain a safe function. \par
	
	Prior studies have only addressed indicators and changes in HRCA signal features at the UES opening and closure moments or during the passage of the bolus through UES, but non of them offered a direct way to detect the DUESO during swallowing. Some of these studies reported the presence of localized maxima of some HRCA signal features at UES opening and closure times \cite{kurosu_detection_2019, perlman_bolus_2005}. One study also observed changes in the acoustic component of HRCA signals while the bolus passed through the UES \cite{moriniere_origin_2008}. Although these studies were essential for establishing the association between UES opening and HRCA signals, they were just descriptive analyses about the patterns in signal features at certain points of time when physiological events occurred. Therefore, in this study we aimed to explore a more advanced predictive profile to detect the DUESO from HRCA signal through considering the time dependency along the swallowing segment. As such we have demonstrated the system’s feasibility on detecting DUESO without VFSS image verification.\par
	
	One major disadvantage of human ratings is the subjectivity which creates an inter-rater tolerance of 82 msec ($\approx\ \pm\ 2.48$ frames at 30 FPS) as reported for measuring swallowing kinematic events \cite{lof_test-retest_1990}. Human ratings of swallow kinematic events can also drift over time and necessitates that raters maintain ongoing intra and inter-reliability over time to maintain an appropriate error tolerance. Having an automated system that is capable of rating the swallowing kinematic events with a comparable human rater accuracy and impregnable to changes over time, is advantageous for swallowing analysis when imaging technology is unavailable, not feasible, or otherwise impractical for evaluating swallowing physiology. Based on the results, we can clearly see that the proposed system accurately detected up to 93.6\% of the actual DUESO with low rates of false positives and negatives occurring only at the borders of DUESO as shown in Fig. \ref{fig:batch_meas} (b). These results were also achieved regardless of gender, age, or diagnosis of the subjects which assures the wide applicability of the system. \par
	
	The system also showed robust performance when applied to a completely independent set of swallows that were collected from a different group of participants with different conditions and never seen in the training dataset. In terms of global measurements, the system achieved a close testing accuracy compared to the validation done through the folds of the original dataset (0.888 vs. 0.9035) and the same for sensitivity and specificity. It didn't come short either on the side of temporal properties of the DUESO, where it captured the UES opening and closure within a 100 msec tolerance in most of the swallows in the independent test set. This confirms that the high quality of DUESO detection can be carried over to completely unseen data and assures a high degree of generalization in the proposed system. \par
	
	It is important to bear in mind that the accuracy of any physiological event detector cannot be judged only through comparison with human ratings which are subject to error too. The sub-events occurring during or after the detected event and their importance to the whole physiological process, control the limits to which the system can be considered accurate because one doesn't want to detect an event with 50 msec accuracy to look for another sub-event that happens within 10 msec of the original event. Previous studies have shown that the important UES events happen slightly after the initial UES opening \cite{cook_opening_1989}. For example, in general, entry of the bolus head into the sphincter defines UES opening; however, in 20\% of swallows, air precedes entry of the bolus by 30-60 msec \cite{cook_opening_1989}. Maximal values of A-P UES diameter were found also to be reached after 70-170 msec of UES opening, depending on the bolus size and other factors \cite{cook_opening_1989}. So, it could be argued that a delayed detection of UES opening is not completely inaccurate if it happens within 100 msec ($\approx$ 3 frames at 30 FPS) after the actual opening. Conversely, anatomic abnormalities leading to reduced DUESO (e.g. cricopharyngeal bar, Zenker diverticulum, hypopharyngeal lesions) would be completely undetectable without imaging leading to the need for further research to determine if HRCA can classify patterns of DUESO that indicate the need for imaging to rule out an anatomic diagnosis reducing DUESO.
	
	In Summary, this study along with others, demonstrates advancements in HRCA signal processing and provides substantial evidence that HRCA signals predominantly reflect the patterns in DUESO and combined with our overall growing research portfolio, swallowing physiological activity. These advancements show the capability of HRCA to provide insight into diagnostic physiological aspects of swallow function and push towards the development of more accessible tools for dysphagia screening within clinical settings. Future research directions for this study include enhancing the detection quality of DUESO while reducing the error between the predicted and actual DUESO and investigating whether characteristic differences in HRCA signal signatures may reflect underlying anatomic or other etiologic explanations warranting investigation with imaging. This point is crucial in that some causes of dysphagia are indeed anatomically based, however in situations in which such diagnoses are suspected and imaging is not available immediately, HRCA certainly shows promise toward providing interim information that can guide management.  Further, the scope of the study will be expanded to include the detection of maximal A-P UES diameter and its time of occurrence solely from HRCA signals.\par

\section{Conclusion}
	In this paper, we proposed an ambitious deep architecture for the temporal identification of the DUESO during swallows by using HRCA signals. Swallows from 116 patients were collected under a standard clinical procedure for different swallowing tasks and materials. 3D acceleration signals of full length swallows, were denoised and fed into a network composed of a two-layer CNN, a 3-layer GRU-based RNN, and 3 fully connected layers to generate the temporal mask marking the time of UES opening and closure during swallows. The proposed system yielded an average accuracy of more than 90\% of the swallow width and more than 91\% of the DUESO width (sensitivity) with a low false positive rate. Moreover, the system showed nearly identical performance when used on an independent testing set from an ongoing clinical trial. Our results have provided substantial evidence that HRCA signals combined with a deep network architecture can be used to demarcate important physiological events that occur during swallowing.

\IEEEtriggeratref{56}



\end{document}